\documentclass[fleqn,10pt]{wlscirep}
\usepackage{amssymb}
\usepackage{amsmath}
\usepackage{graphicx}
\usepackage{epstopdf}
\usepackage{subfigure}
\usepackage{epsfig}
\usepackage{amsfonts}
\usepackage{mathrsfs}
\usepackage[toc,page,title,titletoc,header]{appendix}
\usepackage{dsfont,amsthm,amsbsy}

\def\tr{{\rm tr}}

\title{Investigating disordered many-body system with entanglement in momentum space}

\author[1,+]{Bing-Tian Ye}
\author[1,+]{Zhao-Yu Han}
\author[1,*]{Liang-Zhu Mu}
\author[2,3,4,$\dagger$]{Heng Fan}
\affil[1]{School of Physics, Peking University, Beijing 100871, China}
\affil[2]{Institute of Physics, Chinese Academy of Sciences, Beijing 100190, China}
\affil[3]{School of Physical Sciences, University of Chinese Academy of Sciences, Beijing 100190, China}
\affil[4]{Collaborative Innovation Center of Quantum Matter, Beijing 100190, China}
\affil[*]{muliangzhu@pku.edu.cn}
\affil[$\dagger$]{hfan@iphy.ac.cn}

\affil[+]{these authors contributed equally to this work}

\begin{abstract}
We study the entanglement in momentum space of { the ground state of} a disordered one-dimensional fermion lattice model with attractive interaction. We observe two components in the entanglement spectrum, one of which is related to paired-fermion entanglement and contributes to the long-range correlation in position space.{ The vanishing point of it indicates the localization phenomenon in the ground state of this model.} Additionally, by method of entanglement spectrum, we provide a new evidence to show the transition of two phases induced by interaction, and find {that} this phase transition is not influenced by the disorder. Our result show key characteristics in entanglement for different phases in the system, and provide a novel perspective to understand {localization phenomena.}
\end{abstract}
\begin{document}
\flushbottom
\maketitle
\thispagestyle{empty}

\section*{Introduction}
It is known that a vanishing density of states at fermi level leads to an insulator
in band theory of condensed matter physics.
However, the well-known Anderson localization induced by disorder for noninteracting particles
shows a different mechanism of insulation \cite{Anderson,Evers}.
Recently, there is a { revived} interest in localization due to disorder in the presence of interaction,
known as many-body localization (MBL) \cite{Fleishman,Altshuler,Oganesyan,Basko}.%D. M. Basko, I. L. Aleiner, and B. L. Altshuler, Ann. Phys. (2006)
The MBL may protect a closed quantum many-body system from thermalization which questions our
fundamental assumption in statistical physics, suggesting some emergent conservation laws
for those localized systems \cite{Eisert1}. The localized phase can be closely related with concepts like
eigenstate thermalization hypothesis \cite{Deutsch,Srednicki},
the area-law of entanglement entropy \cite{Plenio,Zanardi,Eisert,Devakul}.
The systems showing MBL phenomenon may have a transition between localized and delocalized phases depending on disorder strength in the
Hamiltonian \cite{Imbrie,Nandkishore}, and many breakthroughs have been made \cite{Anleiner,Vosk,Serbyn}.

Quantum entanglement, a unique feature in quantum physics,
has been shown to be a powerful tool in illustrating the quantum characteristics of
many-body systems. For a pure bipartite state, entanglement spectrum (ES) {contains}
more information of the quantum entanglement than the entanglement entropy which is another measure of entanglement \cite{LiHaldane}.
The ES {is defined as} the logarithm of eigenvalues of reduced density matrix of
a subsystem for a many-body quantum state partitioned into two parts. By investigating the ES in position space, the method has been successfully applied to investigate fractional quantum Hall states \cite{LiHaldane,Rodrguez}, complex paired superfluids \cite{Dubail}, and spin-orbit coupled superconductors \cite{Borchmann}, etc. Also, ES in momentum space \cite{Thomale1,Thomale2,Mondragon,Andrade,Liu1,Liu2,Liu3} and angular momentum space \cite{Bernevig2} can help to
explore new phases of quantum matters of many-body systems such as Chern insulators.  { In these cases, only the ES of ground states is studied, which still indicates the property of excitations in the system.}
Closely related with ES, the entanglement can also be quantified
by the R\'enyi entropy of the reduced density matrix, which can characterize the local convertibility of the bipartite quantum states and {can be used to study quantum phase transition}\cite{Cui}.

{ Some previous work investigated the Anderson localization from the view of the entanglement in the ground states. They concluded that the ES and entanglement entropy in the momentum space can characterize the localized and delocalized phases in Anderson localization transitions\cite{Mondragon,Andrade}.} In this paper, we focus on a fermion lattice model with nearest-neighbor interaction and on-site disorder.
This system is known to have MBL phase and some other quantum phases \cite{Schmitteckert1,Schmitteckert2,Zhao,Berkovits}.
The entanglement in position space has been investigated in spin-$1/2$ Heisenberg model, which is equivalent to a special case of this fermion model, and shows a phase transition from delocalized to localized phases \cite{Yang,Geraedts}.
Different phases come up depending on strength of interaction and disorder. However,
there is uncertainty in the phase transition critical point, and further {charaterization is} necessary.
Here, our work will concentrate on the entanglement in momentum space to analyze {the ground state of} this fermion model. When investigating the entanglement between particles with positive and negative momentum, we depict the features and intuitive picture of two components of it: one component related to the entanglement between particles with opposite momentum would be destroyed by the other component as the strength of disorder grows. {This indicates localizations in the low-lying states, which could also herald the higher-energy MBL phenomenon\cite{Luitz}.} When investigating the entanglement between particles with small and large momentum, {we confirm a phase transition that is not related to localization and originates from the Hamiltonian without disorder}. We also show that the behavior of this transition is robust in the presence of on-site disorder. 

\begin{figure}[t]
\includegraphics[width=1\linewidth]{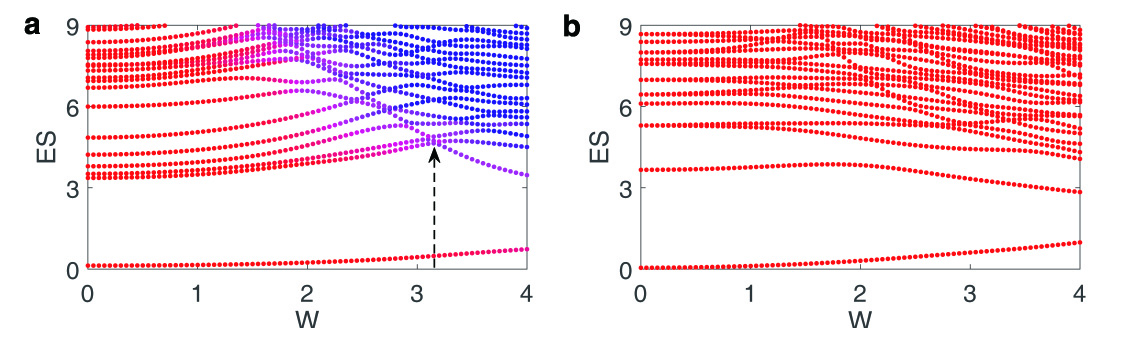}
\caption{\label{ES} ES between the positive and negative momentum parts of the system {with one disorder realization}, of which the remaining particle number $n=5$.  The interaction $U=1$ and the disorder $W$ is the horizontal axis, with the chain length $L=22$ and the total particle number $N=11$. (a) Keep all scattering terms in the Hamiltonian of Eq.~\ref{momentum}, which originate from nearest-neighbor interaction, the color is mapped from relative entropy of coherence, which helps to distinguish paired-particle and chaotic component, and the arrow marks the the point when the chaotic component completely destroys the paired-particle component and thus the critical point; (b) only keep scattering terms originating from interaction between particles with exactly opposite momentum, which are shown in Eq.~\ref{momentumonlypn}, the spectral lines are colored uniformly red.}
\end{figure}

\section*{Results}

\subsection*{The Model}
We consider a spinless one-dimensional fermion model with attractive nearest-neighbor interaction and on-site random disorder. The Hamiltonian of the system is written as,
\begin{equation}
\begin{split}
H=-t\sum^L_{i=1}(&c^{\dag}_i c_{i+1}+H.c.)+\sum^L_{i=1}h_i c^{\dag}_i c_i\\
&+U\sum^L_{i=1}(c^{\dag}_i c_i-\frac{1}{2})(c^{\dag}_{i+1} c_{i+1}-\frac{1}{2}),
\end{split}
\end{equation}
in which $L$ is the number of sites in the chain, the $c^\dag_i$ and $c_i$ are the creation and annihilation operators on site $i$, $U<0$ represents attractive interaction, and $h_i$ randomly but uniformly distributed in $[-W/2,W/2]$ represents on-site disorder. We { let $t=1$ to set the energy scale}.
We remark that when $U=-1$, the model is equivalent to spin-$1/2$ {isotropic} Heisenberg model by applying Jordan-Wigner transformation \cite{JW}. {Moreover, at any value of $U$, the model is equivalent to XXZ model.} 

{We choose the periodic boundary condition, and transform the Hamiltonian into particle number representation in momentum space.} The Hamiltonian in momentum space, up to a constant, is written as,
\begin{equation}
\label{momentum}
\begin{split}
&H=-\frac{2}{\sqrt{L}}\sum_{p}\mathrm{cos}(\frac{2p}{L}\pi)a^{\dag}_p a^{}_p+\sum_{p,q}g_{p-q}a^{\dag}_p a^{}_q\\
&+\frac{4U}{L}\sum_{\substack{p+q=p'+q'\\p>q,p'>q'}}\mathrm{sin}(\frac{p-q}{L}\pi)\mathrm{sin}(\frac{p'-q'}{L}\pi)a^{\dag}_{p'} a^{}_{p} a^{\dag}_{q'} a^{}_{q},
\end{split}
\end{equation}
in which $g_p$ is Fourier transform of  $h_i$, $a^\dag_p$ is creation operator of fermion with pseudo-momentum $2\pi p/L$,
and the summation is performed in the first Brillouin zone.

To obtain the ES, we can calculate the reduced density matrix of the subsystem, i.e. $\rho_A=\tr_B \rho_{AB}$, and the eigenvalues of the reduced density matrix are $\{\lambda_i\}$. In this paper, we use two ways to divide {the} system into two subsystems in the reciprocal lattice and study the corresponding ES: {to investigate the entanglement between particles with positive and negative momentum we divide the lattice into left and right parts}, { and to study the entanglement between small and large momentum we divide the reciprocal lattice into inner and outer parts}. 

Since the total particle number operator commutes with the Hamiltonian, we can focus on the ground state in the subspace of certain particle number $N$. We remark that $N$ should be an odd number to avoid having a degenerate Fermi sea in calculation\cite{Thomale2} {(in the sense of average when there is disorder)}. Furthermore, when we divide the system, the reduced density matrix of its subsystem must be in block-diagonal form, and every block corresponds to a number $n$ representing the amount of remaining particles in the subsystem {(see Methods)}. Thus, we can use $n$ as a parameter to mark each spectral line in ES.

\begin{figure}[t]
\includegraphics[width=\linewidth]{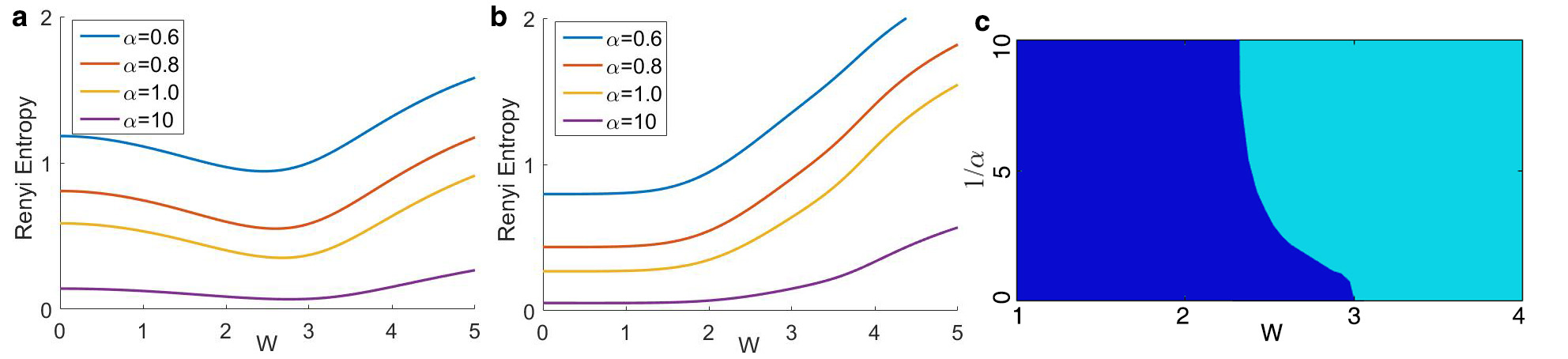}
\caption{\label{EE}The behavior of REE between the positive and negative momentum parts of the system {(with the same disorder realization as in Fig 1.)}, of which the remaining particle number $n=5$. The interaction $U=1$ and the disorder $W$ is the horizontal axis, with the chain length $L=22$ and the total particle number $N=11$. Renyi entropy for different $\alpha$ when (a) {Keeping all scattering terms in the Hamiltonian of Eq.~\ref{momentum}, which originate from nearest-neighbor interaction}, (b) only keeping scattering terms originating from interaction between particles with exactly opposite momentum, { which are shown in Eq.~\ref{momentumonlypn}}. (c) The sign distribution of $\partial_W S_\alpha$ on the $1/\alpha-W$ plane. $\partial_W S_\alpha$ is positive in cyan (light gray) regions and negative in the blue (dark gray) regions. When the sign takes on both positive and negative at certain $W$, paired-particle component of entanglement still exists. Otherwise, it is completely destroyed.}
\end{figure}

\subsection*{Entanglement of positive-negative momentum}

We first investigate the ES by dividing the system into positive and negative momentum parts. We primarily concentrate on { the} ES in the sector corresponding to {one where} the remaining particles in positive momentum parts equals to a half of total particles (except the zero-momentum particle). For a fixed $U$, the ES  {at different} disorder strengths, $W$'s, are shown in Fig.1a. Apart from the lowest one spectral line corresponding to the ground state without interaction or disorder, two components can be seen in ES. Based on the fact that the states corresponding to the spectral lines have different coherence for different component, it helps us distinguish the two components to map the colors of each line from relative entropy of coherence, which can quantify coherence of the states \cite{Baumgratz}. One component, named paired-particle component, is the main ES when $W$ is small. As $W$ increases, the other component, named chaotic component, goes lower, showing that it becomes more important. The descending chaotic component crosses with the original second lowest spectral line at the point marked by an arrow in Fig.1a, showing that the chaotic component completely destroys the paired-particle component.

We also study the R\'enyi entanglement entropy (REE) in the same sector. It is defined as $S_{\alpha}=\frac{1}{1-\alpha}\mathrm{log}\sum_i \lambda_i^\alpha$, in which $\alpha$ is a positive parameter { and $\lambda_i$ are the eigenvalues of the reduced density matrix}. As shown in Fig.2a, the REE reaches a minimum for all $\alpha$, which indicates the relative importance of different entanglement components changes.  {To better illustrate this picture}, we investigate the sign of $\partial_W S_\alpha$, which shows how the strength of entanglement varies as $W$ changes, as shown in Fig.2c. When fixing W to be smaller than a certain value, the change of the signs of $\partial_W S_\alpha$ for different $\alpha$'s shows there is no global shift of all $S_{\alpha}$. Since different $\alpha$'s in R\'enyi entropy represents different scales to measure entanglement, two directions of shift in $S_{\alpha}$ with different $\alpha$'s mean there are two components of entanglement and {they change} in different ways as $W$ increases. Therefore, considering that the chaotic component becomes more important as $W$ increases, only when all $\partial_W S_\alpha$ have the same sign at a certain $W$, paired-particle component is completely destroyed.
This value of $W$, which will be identified later, indicates the critical point of a phase transition \cite{Cui}, {the other aspects of which have been manifested in some previous works on the ground state of this model\cite{Berkovits,Schmitteckert1,Zhao}.}

{We now show that the paired-particle} component indeed originates from entanglement between particle pairs with opposite momentum, and is related with long-range correlation. It is known that the attractive interaction may cause fermions to pair, and the interaction terms relating fermions with opposite momentum plays a central role in this system. Similar to BCS theory of superconductivity, if we only keep those terms relating particles with exact opposite momentum and drop other scattering terms, then perform Bogoliubov transformation, we can acquire a set of newly defined quasi-particles which can be seen as the linear combination of the original fermions with opposite momentum \cite{Bardeen,Kitaev}. Hence the eigenstate of the system can be solved in the single-particle picture of these quasi-particles. The ground state, as well as excited states, is exactly a basis of particle number representation of quasi-particles. These pairs represent the entanglement between particles with opposite momentum, which induces long-range correlation. Also, the pairs, which could not be destroyed when adding disorder (only with exactly opposite momentum scattering terms), would contribute to p-wave superconductivity \cite{Alicea,Berkovits}.

According to the above consideration, the kept interaction term could be written as,
\begin{equation}
\label{momentumonlypn}
V=\frac{4U}{L}\sum_{p,p'>0}\mathrm{sin}(\frac{2p}{L}\pi)\mathrm{sin}(\frac{2p'}{L}\pi)a^{\dag}_{p'} a^{}_p a^{\dag}_{-p'} a^{}_{-p},
\end{equation}
Since ES can reflect entanglement in detail, the ES of the ground state of the Hamiltonian with this interaction can be used as a signature of the long-range correlation. We investigate this ES as shown in Fig.1b. The absence of turning point in the second lowest spectral line in Fig.1b suggests that no chaotic component destroy the original component in ES relating to pair-particle entanglement. Furthermore, the absence of the chaotic component in this case demonstrates that it is generated by the scattering term between particles with inexactly opposite momentum. Also, REE reflects the same picture. As shown in Fig.2b, $S_{\alpha}$ monotonically increases for all $\alpha$, which indicates that there is only paired-particle component of entanglement when only reserving interaction term between opposite momentum. 

\begin{figure}[t]
\centering
\includegraphics[width=0.8\linewidth]{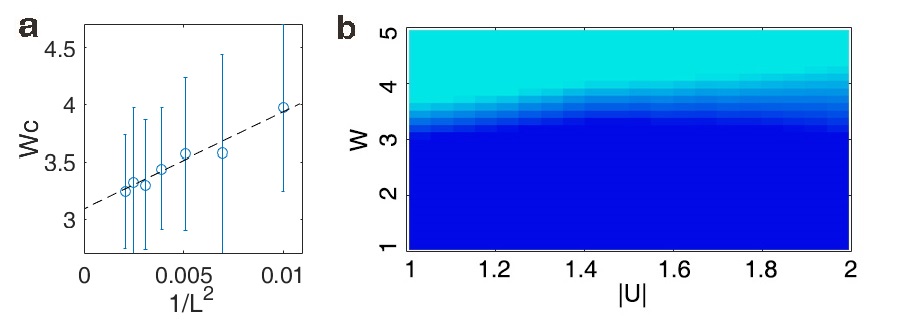}
\caption{\label{phase} {(a) The fitting relation between $1/L^2$ and the critical $W_c$. Every data point is based on results on 15 disorder realizations at each chain length.} (b) The phase diagram when the interaction $|U|\in [1,2)$ with the chain length $L=20$ and the total particle number $N=10$. The cyan (light gray) region is the localized phase; the blue (dark gray) region is the delocalized phase. { The uncertainty due to finite size with random disorder is estimated by the standard deviation on the results of 15 disorder realizations and shown by gradient colors.}}
\end{figure}

Thus, the physical picture is clear. When disorder is weak, the paired-particle component plays the main role, suggesting that the system is delocalized. As the disorder becomes stronger, although the disorder does not destroy paired-particle component directly, it makes the chaotic component more important and decisive, which means the long-range correlation is destroyed and the system becomes localized. We have already seen in Fig.1a that as the chaotic component goes lower, the position of the second lowest line would meet a turning point when the paired-particle component loses its leading role. This point, along with the vanishing of {blue} region in \ref{EE}c which occurs slightly earlier than this point due to the effect of the lowest spectral line, indicates the transition from delocalized to localized phase. Since this picture is based on the modes of quasi-particle which are particle pairs, and does not depend on whether these modes are excited, {we hope that it could also apply to the first few excited states of the system and thus could provide evidences for the interpretation of the higher-energy MBL phenomenon.} As is analyzed above, by calculating the ES according to different $W$, we can find the turning point of second lowest spectral line and use it to mark the critical point. At $U=-1$, {we did the statistics of 30 samples for this point and found that this transition occurs averagely at $W_c=3.2$ and the standard deviation is $0.5$}. {We also investigated the finite size scaling behavior of the critical point. As shown in Fig.3a, we see the critical point indeed converges at large chain length, to a value of approximately $3.1$. This result is consistent with previous study on the MBL edge\cite{Luitz}, and is slightly smaller than the MBL transition point\cite{Yang,Pal}.} When $U\in [-1,-2)$, we find the same picture still applies, so we investigated 15 samples and acquire the phase diagram of this region as shown in Fig.3b.

\begin{figure}[t]
\includegraphics[width=\linewidth]{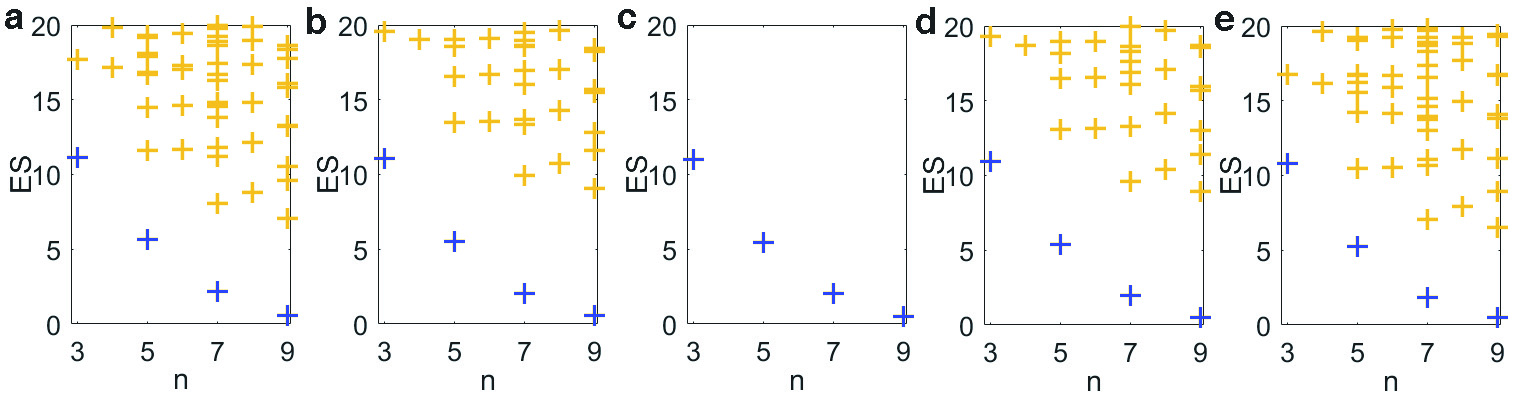}
\caption{\label{ES3}ES between small and large momentum parts of the system, when there is no disorder. The chain length $L=22$ and total particle number $N=11$. The blue (dark gray) lines represent the lowest spectrum, and the orange (light gray) lines represent the upper parts. The strength of interaction $U$ equals to (a)$1.97$, (b)$1.99$, (c)$2.00$, (d)$2.01$, (e)$2.03$.}
\end{figure}

\subsection*{Entanglement of small-large momentum}
In this case, we first investigate ES of the system without disorder. We divide system into small and large momentum parts along the fermi level of non-interacting case. When adding interaction, the particles would be scattered to above the fermi level, leaving holes below. Thus, the entanglement between these two parts can be seen as entanglement between particles and holes. As is shown in Fig.4, in each sector of which remaining particle number $n$ equals to total particle number $N$ minus an even number, there is a gap between the lowest spectral line and upper parts. As $|U|$ increases, the gap increases to reach a sharp maximum when $U=-2$, and then decreases. Since the apparent entanglement between different sectors is due to that the particle number is a good quantum number and thus does not fundamentally reflects the entanglement between particles and holes, the only one spectral line in every individual sector shows the vanishing of their entanglement, which suggests a transition from one structure of entanglement to another and thus resulting in a probable phase transition. This is consistent with the conclusion achieved by typical other methods \cite{Yang1,Yang2}. 

When there is disorder but not too large, the gap in the {ES as a function of $U$} is shown in Fig.5. Although the entanglement between particles and holes does not completely vanish, it gets to be the weakest and thus the gap still reaches the maximum {near $U=-2$}, which shows {that the critical point is insensitive to the disorder}\cite{Schmitteckert2,Berkovits}.

\begin{figure}[t]
\centering
\includegraphics[width=0.5\linewidth]{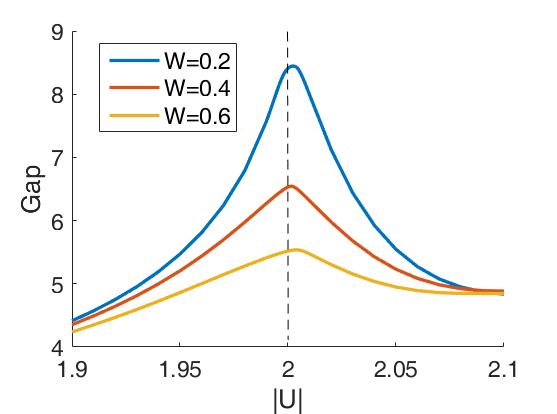}\\
\caption{\label{gap}The gap in the ES, of which the remaining particle number $n=9$. The chain length $L=22$ and the total particle number $N=11$. { We use the same disorder realization as in Fig 1.} Different colors (gray levels) represent different disorder strength $W$.}
\end{figure}

\section*{Discussion}

By partitioning the reciprocal lattice into two subsystems in different ways, we investigate the entanglement in momentum space of a fermion lattice model with interaction and disorder. When partitioning particles with positive and negative momentum, we find that the ES consists of two components. Comparing with the ES of system possessing interaction only between particles with opposite momentum, we show that one of the component {originates} from paired-particle entanglement, i.e. the entanglement between particles with opposite momentum, and this entanglement represents long-range correlation in position space. However, the other component, chaotic entanglement, may {destroy} the paired-particle entanglement and induce { a localization phenomenon in the ground state.} Based on this picture, we obtain the critical point between the delocalized phase and localized phase of this model, and found it close to the critical point given by previous studies. When dividing the momentum space into small and large momentum parts, 
we observe that the behavior of the gap in ES can be used to identify a phase transition in this system, and can show that this transition is {insensitive} to the disorder.

Our work provides a clear picture in understanding the characteristic of disordered fermion lattice models and  the nature of related localization transitions from a novel perspective - by investigating entanglement in momentum space. The methods including entanglement entropy, entanglement spectrum, R\'enyi entanglement entropy and its derivative can also be {applied } to other systems in studying similar phenomena.

\section*{Methods}
{As for the ground state of the system, the wave function can then be Schmidt decomposed in the form $|\Psi\rangle=\sum_{i}\sqrt{\lambda_{i}}|A_i\rangle|B_i\rangle$ \cite{Schmidt,Nielson}, where $\{ |A_i\rangle\}$ and $\{ |B_i\rangle\}$ are two sets of orthonormal basis vectors of two subsystems A and B respectively, and $\{ -\ln\lambda_i\}$ are the so-called ES.}
 
Considering the long-range interaction in the reciprocal lattice, the DMRG method may not be suitable for solving the ground state of this model in particle number representation of momentum space \cite{Schollwock}, so we use Arnoldi method to acquire it \cite{Arnoldi,Arnoldi1}. To obtain the ES, we can calculate the reduced density matrix of the subsystem, i.e. $\rho_A=\tr_B \rho_{AB}$, and find the eigenvalues of the reduced density matrix, which are $\{\lambda_i\}$. We will also remark that since the reduced density matrix of the subsystem is in block-diagonal form corresponding to different remaining particle numbers $n$ in the subsystem, the diagonalizing processes can be restricted in each block.

As below we prove that the reduced density matrix of the subsystem in this case must be in block-diagonal form, and every block corresponds to a number $n$ representing the amount of remaining particles in the subsystem. This result can also be generalized in the cases when the systems have any local conserved quantities. We fist choose two sets of basis $\{ |A_{n,i}\rangle\}$ and $\{ |B_{n',j}\rangle\}$, in which $n$ and $n'$ represent the particle numbers in subsystem A and B respectively, and $i,j$ are some other quantum numbers.  Then the state of the system can be written as
\begin{equation}
\label{proof1}
|\psi\rangle =\sum_{n,n',i,i'}\lambda_{n,n',i,i'}|A_{n,i}\rangle |B_{n',i'}\rangle. 
\end{equation}
Since $|\psi\rangle$ is the eigenstate of the total particle number with the eigenvalue of $N$, we have $n+n'=N$ for all $\lambda_{n,n',i,i'} \neq 0$. Thus, the reduced density matrix $\rho_A$ of subsystem A is
\begin{equation}
\label{proof2}
\begin{split}
\rho_{n,i;m,j} &=\sum_{n',i'}\lambda_{n,n',i,i'}\lambda^*_{m,n',j,i'}\\
&=\sum_{n',i'}\delta_{n,N-n'}\delta_{m,N-n'}\lambda_{n,n',i,i'}\lambda^*_{m,n',j,i'}\\
&=\sum_{n,i'}\delta_{n,m}\lambda_{n,N-n,i,i'}\lambda^*_{m,N-m,j,i'}, 
\end{split}
\end{equation}
which is a block-diagonal matrix, and $n$ can be used to distinguish different blocks.

\bibliography{MBL}

\section*{Acknowledgements}

This work is supported by MOST of China (2016YFA0302104, 2016YFA0300600), NSFC (91536108), NFFTBS (J1030310, J1103205), the Strategic Priority Research Program of the Chinese Academy of Sciences (XDB01010000, XDB21030300), the Young Elite Program for Faculty of Universities in Beijing, and Training Program of Innovation for Undergraduates of Beijing.

\section*{Author contributions statement}

B.T.Y. and Z.Y.H. conceived the idea, developed the software, analyzed the results and drafted the manuscript. All authors revised the final manuscript. L.Z.M and F.H. gave constructional advice and supervised this work.

\section*{Additional information}
Competing financial interests: The authors declare no competing financial interests.

\end{document}